\documentclass[aps,pra,groupedaddress,showpacs,twocolumn]{revtex4}

\usepackage{braket}
\usepackage{graphicx}
\usepackage{epstopdf}
\usepackage[latin1]{inputenc}
\usepackage{amsmath,amsbsy,amsfonts,amssymb}
\usepackage{epsfig}
\usepackage{hyperref}
\newcommand{\wn}{\mbox{cm$^{-1}$}}

\begin{document}

\title{Nonlinear Zeeman effect in photoassociation spectra of $^{40}$Ca near the $^3$P$_1$+$^1$S$_0$ asymptote}

\author{Eberhard Tiemann$^1$}

\affiliation{$^1$Institut f\"ur Quantenoptik, Leibniz Universit\"at Hannover, Welfengarten 1, 30167 Hannover, Germany}

\author{Max Kahmann$^{2}$}

\altaffiliation[Current address: ] {TRUMPF Laser- und Systemtechnik GmbH, Johann-Maus-Str. 2, 71254 Ditzingen, Germany}

\author{Evgenij Pachomow$^2$}

\email{evgenij.pachomow@ptb.de}

\author{Fritz Riehle$^2$}

\author{Uwe Sterr$^2$}

\affiliation{$^2$Physikalisch-Technische Bundesanstalt (PTB), Bundesallee 100, 38116 Braunschweig, Germany}

\date{\today}

\begin{abstract}
We present calculations of the Zeeman effect of narrow photoassociation lines of $^{40}$Ca near the $^3$P$_1$ + $^1$S$_0$ asymptote. Using a coupled-channel model we find a nonlinear Zeeman effect that even at low fields of a few mT amounts to several kHz. With this model we analyze previous measurements and give corrected long range dispersion coefficients of the $^3\Pi_{u}$ and $^3\Sigma^+ _{u}$ states.      
\end{abstract}


\maketitle

\section{Introduction}

The scattering properties of alkaline earth atoms like Mg, Ca, Sr and the isoelectronic species Yb are currently subject of intense investigations, both theoretically and experimentally. 
These properties are highly relevant e.g. for the generation of ultra-cold atoms in the quantum degenerate regime \cite{kra09,ste13d,sug13}, the evaluation of collisions that may limit the accuracy of the best optical clocks \cite{ush15, nic15}, quantum computation \cite{yi08}, the determination of molecular potentials with high accuracy \cite{jon06}, or the production of ultra cold molecules \cite{rei12} with the prospect of studying ultra-cold chemistry. Many of these investigations are performed by using the photoassociation process, where a molecule is built from two ground state atoms in a resonant laser field. 
The intercombination transitions $^1$S$_0 - ^3$P$_1$ of alkaline earth atoms with their extremely narrow natural line width of $\Gamma/2\pi$=7~kHz and $\Gamma/2\pi=374(9)$~Hz for Sr and Ca, respectively, present almost ideal model systems to perform such investigations with kilohertz accuracy \cite{mcg13,kah14,kah14a}. 

\begin{figure}
\includegraphics[width=0.9\columnwidth]{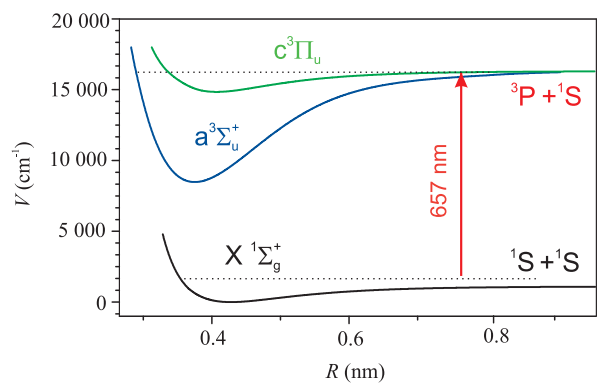}
\caption{Relevant Hund's case (a) molecular potentials of Ca$_2$ for photoassociation near the ${^3P_1}-{^1S_0}$ asymptote ($\lambda=657$~nm). The position of the transition is not to scale.}\emph{}
\label{Fig:potentiale}
\end{figure}

Despite of their similar electronic configuration, the twenty times smaller line width of Ca with respect to Sr leads to a striking difference for their long range potentials. In Ca the long range potentials are dominated by the van der Waals ($C_6$) interaction at the asymptote $^3$P + $^1$S. 
Consequently, the small $C_3$ coefficient (dipole-dipole interaction) in Ca results in two excited attractive potentials c$^3\Pi_{u}$ and a$^3\Sigma^+ _{u}$ (Fig. \ref{Fig:potentiale} and Eq. \ref{eq:potential0}) which have a similar $1/R^6$ dependency as the ground state potential \cite{all03}. 
The c potential is purely attractive while the a potential with its $\Omega = 1$ component has a maximum of only $h\times18~$MHz at a distance of $R=5.5$~nm due to the repulsive $C_3$ term. The height of this maximum is small compared to the binding energy of the least bound state ($\approx 983$~MHz) which is still dominated by the $C_6$ coefficient.
 
It was shown \cite{kah14} that the molecular states within these asymptotically similar potentials can couple by spin-orbit and rotational interaction remarkably even for the lowest angular momentum $J=1$. This behavior leads to a striking dependence of the molecular g factor on the rovibrational levels. 
The quadratic Zeeman effect that is significant in the photoassociation spectra of $^{88}$Sr \cite{mcg13} is much smaller in Ca because of the different potential structure in Ca$_2$ and Sr$_2$ by the $C_6$ and $C_3$ terms, respectively. This results in a more widely spaced level structure for Ca$_2$ at the asymptote, whereas for Sr$_2$ the levels of the $\Omega = 0+$ state (the plus sign referring to the total parity)are so close to the asymptote that the quadratic Zeeman effect is dominated by the coupling to the continuum above the asymptote. Thus the quadratic Zeeman contributions are significantly different in both molecules. 
In \cite{kah14} it was therefore treated as being negligible under the experimental conditions and only considered in the uncertainty budget of the derived experimental g factors. 
In the present paper we give a more elaborate treatment and calculate the Zeeman effect of the narrow photoassociation lines of $^{40}$Ca near the $^3$P$_1$ + $^1$S$_0$ asymptote. 
Correcting previous measurements \cite{kah14} we derive improved values of the $C_6$ and $C_8$ coefficients and g factors \cite{kah14a}.  

\section{Zeeman effect of an atomic pair $^3$P +$^1$S}
The model is based on Hund's case (a) potentials for the states $a^3\Sigma^+_u$ and $c^3\Pi_u$ (see Fig. \ref{Fig:potentiale}). The short range potentials \cite{all05} are extended for $R > 2$~nm as 
\begin{eqnarray}
\label{eq:potential0}
V_{a}=& -\frac{-4C_{3}^{(0)}}{R^3} -\frac{C_{6}^a}{R^6}-\frac{C_{8}^a}{R^8}  \\
V_{c}=& -\frac{2C_{3}^{(0)}}{R^3}-\frac{C_{6}^c}{R^6}-\frac{C_{8}^c}{R^8} \nonumber.
\label{eq:potential1}
\end{eqnarray}

In \cite{kah14} we calculated the molecular g factor from the expectation values of the Zeeman interaction of eigenstates of the field-free Hamiltonian; this corresponds to the first-order approximation. 
Here, we extend this theoretical approach by diagonalization of the full Hamiltonian being the sum of the field-free part and the Zeeman part. The first one is described in our previous work and the later one is set up as:
\begin{equation}
\begin{split}
H_\mathrm{Z}=\mu_Bg_jB_0j_z ,
\label{eq:Zeeman}
\end{split}
\end{equation}
where $\mu_B$ is the Bohr magneton and $g_j$ the atomic g factor of state $^3$P$_j$ and $j_z$ is the projection of the total atomic angular momentum $j$ onto the space fixed axis parallel to the external magnetic field with magnitude $B_0$. The actual Zeeman energy for magnetic fields below 10~mT is much smaller than the spin-orbit energy in Ca, thus recoupling of $j$ to separate terms of orbital and spin angular momenta is unimportant.
The atomic state $^1$S$_0$ does not contribute to the Zeeman energy. We neglect the Zeeman energy in the excited state by the pair rotation with the angular momentum $l$, for which the effective magnetic moment is only in the order of the nuclear magneton, and could give rise to shifts of not more than   
1~kHz for 100~$\mu$T field strength. 
 
The calculation is done in a Hund's case (e) representation $\ket{(L,S)j,l,J,M}$. The first three quantum numbers denote the atom  $^3$P$_j$, the constant quantum numbers for state $^1$S$_0$ are suppressed as well as $(LS)$ in the state vector below because of no importance here. $J$ and $M$ are the total angular momentum and its projection on the space fixed axis. The matrix elements of operator $H_\mathrm{Z}$ (Eq. \ref{eq:Zeeman}) are calculated with conventional angular momentum algebra
\begin{equation}
\begin{split}
\bra{j',l',J',M} j_z\ket{j,l,J,M} \\
=\delta_{l',l}\delta_{j',j}(-1)^{j+l+J'+J+1-M}\\
 \cdot \mathrm{w3j}(J',1,J,-M,0,M)\cdot \mathrm{w6j}(j',J',l,J,j,1)\\
\cdot [(2J+1)(2J'+1)(2j+1)(j+1)j]^{1/2}
\label{eq:Zeeman_mat}
\end{split}
\end{equation}
w3j and w6j are Wigner's 3j- and 6j-symbols, respectively. The operator has non-vanishing matrix elements for $\Delta J= 0,\pm1$ only. In the present experiments only low magnetic fields were applied, thus we truncate the matrix and span $J \leq 3$ for calculating the observed states, which have still fairly good quantum numbers $J\approx 1$.

\begin{table*}[t]
\centering
\begin{tabular}{|c|c|c|c|c|c|c|c|c|c|c|}
\hline
\multicolumn{2}{|c|}{~level~} & $\Delta_{\rm b}^{\rm calc}$ & $\Delta_{\rm b}^{\rm exp}$ & $\delta$ & $g^{\rm calc}$ & $|g^{\rm exp}|$  & $\chi^\mathrm{scal}$ & $\chi^\mathrm{tens}$ & $\Delta_{\rm Zeeman}^{\rm nl}$ \\

$~v'$ & $\Omega$ & GHz & GHz & kHz &  &  & 10$^{10}\cdot$Hz/T$^2$ & 10$^{10}\cdot$Hz/T$^2$ & kHz \cite{kah14a}\\

\hline
-1 & 0 & -0.308684 & ~-0.308700~(8) & -16 & -0.274 & 0.276~(2)  & 72.6(3) & -0.7(3)  & +30 \\

\hline
-1 & 1 & -0.982984 & ~-0.982977~(7) &  7 & ~1.069 & 1.074~(4)  & -59.3(3) & -8.5(3)   & -17 \\

\hline
-2 & 0 & -4.649238 & ~-4.649209~(22) & 29 & -0.147 & 0.147~(6) & 18.7(3) & 2.6(4)  & +5 \\

\hline
-2 & 1 & -7.411921 & ~-7.411933~(13) & -12 & ~0.902 & 0.901~(29)  & -35.2(3) & -0.9(4)  & -14 \\

\hline
-3 & 0 & -17.857276 & -17.857276~(8) & 0 & -0.079 & 0.080~(6) & 7.3(3) & 1.3(4)   & +2 \\

\hline
-3 & 1 & -24.539436 & -24.539435~(8) & 1 & ~0.831 & 0.838~(26)  & -25.4(3) & 0.5(4)  & -11 \\

\hline
-4 & 0 & -44.758207  &  & &  -0.0435 &   &  & & \\
\hline
-4 & 1 & -57.852473  &  & & ~0.7946 & &  & & \\
\hline
\end{tabular} 
\caption{Measured and calculated binding energies $h \Delta_{\rm b}$ (with $J=1$) at zero field, their difference $\delta$ and molecular g factors of the $c0_u^+$ and $(a,c)1 _u$ states for vibrational level $v'$ counted from the asymptote. 
Results of our calculation: Quadratic Zeeman effect described by scalar and tensorial susceptibility 
$\chi^\mathrm{scal}$ and $\chi^\mathrm{tens}$, frequency shift $\Delta_{\rm{Zeeman}}^{\rm{nl}}$ of the $M=0$ component, calculated for a magnetic field of $B=0.285(7)$ mT.}
\label{tab:measurements}
\end{table*} 

\begin{figure}
\includegraphics[width=0.97\columnwidth]{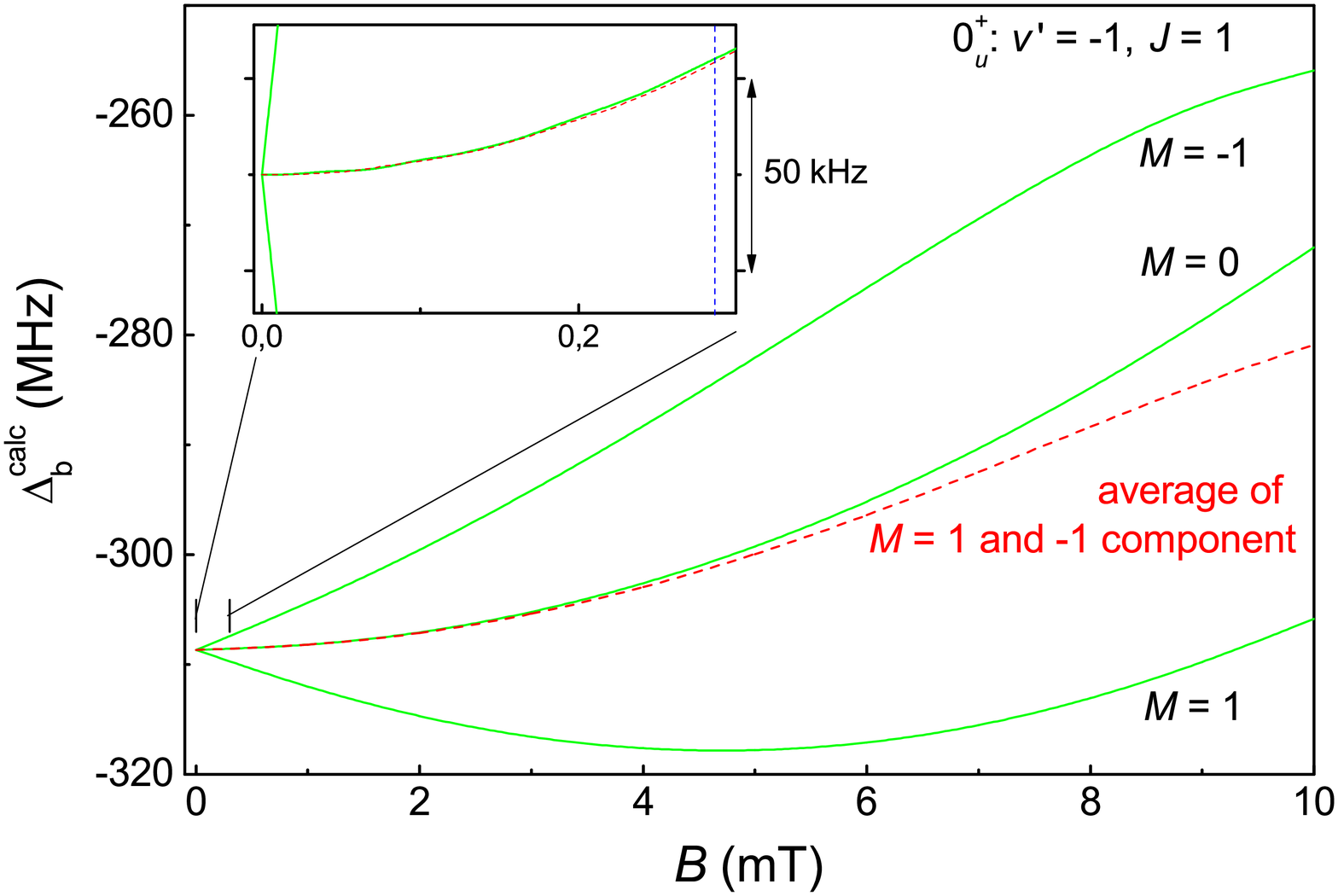}
\includegraphics[width=0.97\columnwidth]{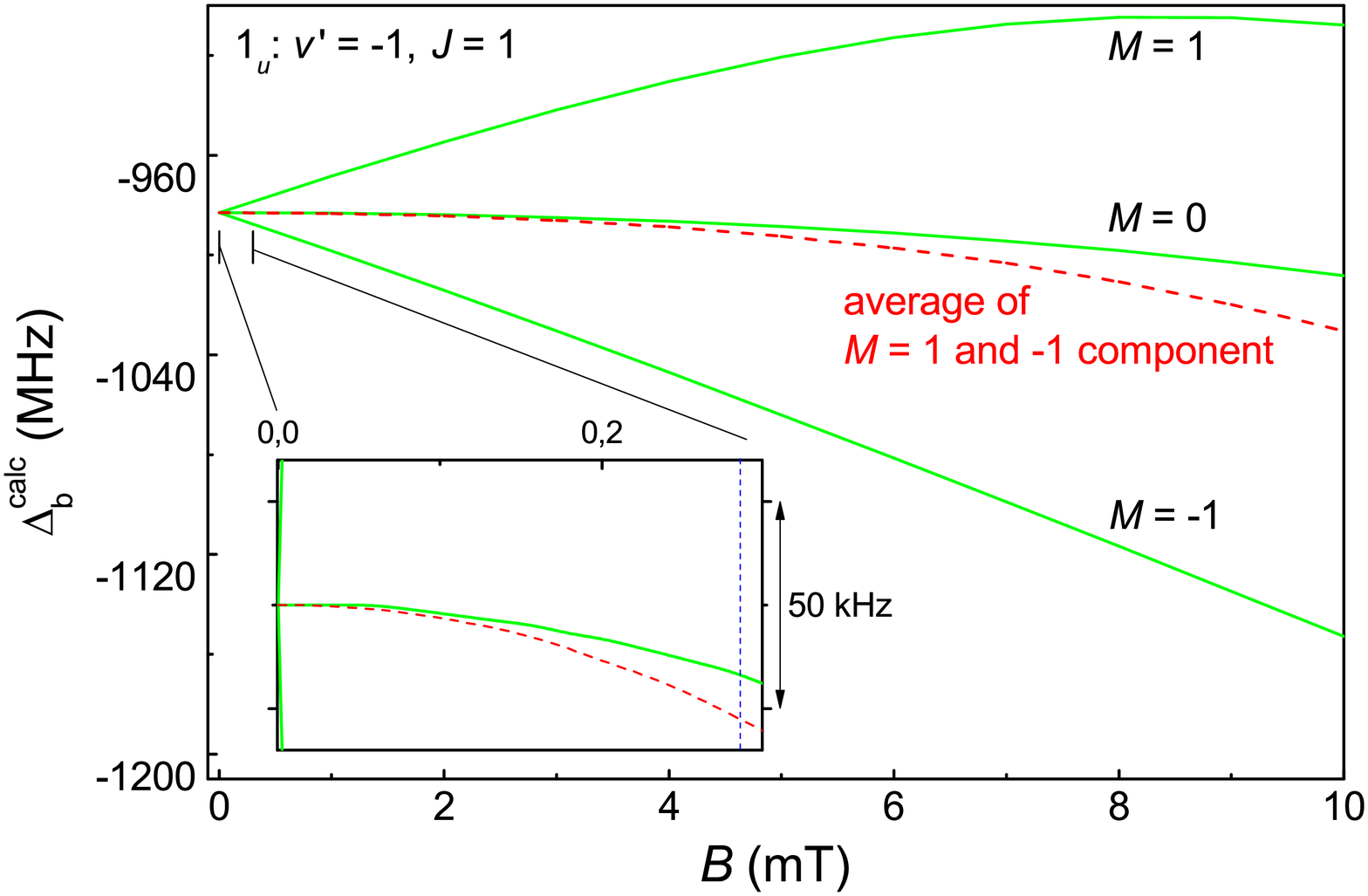}
\caption{Binding energies $h \cdot  \Delta_b$ of the \textit{v'} = 1 lines in the 0 and 1 potential. The red dashed line indicates the average of the $M = +1$ and $M=-1$ energy. The dashed vertical line in the insets  indicates the magnetic field of 285~$\mu$T used in the experiment \cite{kah14}. In the insets the $M=\pm 1$ components are barely visible on the left side, because the linear Zeeman effect is huge compared to the quadratic one.}
\label{Fig:Zeeman_function}
\end{figure}

Fig. \ref{Fig:Zeeman_function} shows the calculated Zeeman energy for the most weakly bound levels of state $0^+_u$ and $1_u$ demonstrating clearly the nonlinear behavior for magnetic fields below 10~mT.
The excited ungerade state c$0^+_u$ and the strongly mixed state (a,c)$1_u$ asymptotically approach the $^3$P$_1 + ^1$S$_0$ state and correlate to the potentials denoted $c^3\Pi_u$ and $a^3\Sigma^+_u$ in Hund's coupling case (a) shown in Fig. \ref{Fig:potentiale}.

The inset in Fig. \ref{Fig:Zeeman_function} gives a zoom at low magnetic field presenting the quadratic Zeeman shift of 30~kHz and -17~kHz for the $M=0$ component of the least bound states in the two levels at the low field of 285~$\mu$T used in the former experiment \cite{kah14}. 
The dashed lines represent the average of the $M=\pm1$ levels thereby canceling their linear Zeeman energy. 
At low magnetic field these averages coincide well with the $M=0$ level, which indicates a small tensorial second-order Zeeman shift for the state $1_u$. 
For comparison, the quadratic Zeeman shift of the atomic line ($^1S_0 - ^3P_1$), mostly resulting from admixture of the other atomic fine structure terms $j=0$ and $j=2$, is 
$6.3 \times 10^7~\mathrm{Hz/T^2}$ \cite{bev98} resulting in a shift of 5~Hz for the actual field and thus
more than a factor of $10^3$ smaller than for the weakly bound molecular states.

The quadratic Zeeman shift was previously not included in the extrapolation of the measured energies to zero field \cite{kah14}. We calculated this shift $\Delta^{nl}_{Zeeman}$ (column 10 of Table \ref{tab:measurements}) for the $M=0$ component of all observed levels for 285~$\mu$T and applied them as corrections to the data in \cite{kah14} for obtaining the zero field frequencies $\Delta^{exp}_p$ (collected in column 4 of Table \ref{tab:measurements}).  The experimental molecular g-factors were derived from the $M=\pm 1$ components correcting also the quadratic Zeeman effect are shown in column 7 of Table \ref{tab:measurements}. Because the experiment did not distinguish between $M=\pm 1$ the sign of the g-factor is not determined.

As in \cite{kah14} the corresponding energies of the observed six PA resonances and their molecular g factors were fitted simultaneously. A non-linear least squares fit routine varies the slope of the short range repulsive branch (to get the proper phase of the long range wave function) and the long range parameters $C_6$ and $C_8$. 
The $C_3^{(0)}$ coefficient was held fixed at the value calculated from the life time of $^3P_1$ state of Ca, see ref. \cite{kah14}. 
The fit results for the energies and the g values are given in columns 3 and 6 of Table \ref{tab:measurements}. The corresponding long range parameters $C_6$ and $C_8$ are listed in Table \ref{tab:Ci} together with theoretically derived values. Within the estimated uncertainty of 5\% and 20\% for the $C_6$ and $C_8$ parameters, respectively, our values agree with the theoretically determined ones.    

\begin{table}[t]
\centering
\begin{tabular}{|c||c|c|c|c||c|c|c|}
\hline
state & \multicolumn{4}{|c||}{  $C_6$  } & \multicolumn{3}{|c|}{ $C_8$}\\
\hline
 & \multicolumn{4}{|c||}{  $10^{7}$ \wn \AA$^6$  } & \multicolumn{3}{|c|}{ $10^{9}$ \wn \AA$^8$}\\
\hline
 & this work & \cite{kah14} & \cite{ciu04}& \cite{mit08} & this work & \cite{kah14} & \cite{mit08}   \\
\hline
 $c^3\Pi_u$   & 1.209 & 1.199 & 1.187 & 1.226 & 0.238 & 0.289 & 0.266  \\
 $a^3\Sigma_u$& 1.335 & 1.353 & 1.313 & 1.358 & 0.905 & 0.818 & 1.057   \\
\hline
\end{tabular}
\caption{Long range parameters derived from the present work at the asymptote $^3$P$_j + ^1$S$_0$ in comparison with theoretical results from Mitroy and Zhang  \cite{mit08} and Ciury{\l}o \textit{et al.} \cite{ciu04}.}
\label{tab:Ci}
\end{table}

From the calculations we can extract the low-field coefficient of the quadratic Zeeman effect that can be written as a sum of a scalar and a tensorial part:
\begin{equation}
	\Delta E^{(2)}_\mathrm{Z}(B, M) = \\
	-\frac{B^2}{2} \left[\chi^\mathrm{scal} + (3M^2-J(J+1)) \chi^\mathrm{tens}\right]
\label{Eq:tensor}
\end{equation}

Results  are presented in Table \ref{tab:measurements}, columns 8 and 9 fitting the field range up to 0.3~mT of calculations as shown in Fig. \ref{Fig:Zeeman_function}. 
It is striking that the scalar part alternates in sign between the states $\Omega=0^+$ and $\Omega=1$. This is directly related to the position of the perturbing states with respect to the observed states, which are all $J=1$ with negative parity (noted below for short by a minus sign behind the $J$ quantum number), i.e. levels with so-called {\it e} symmetry. From eq. \ref{eq:Zeeman_mat} the non-diagonal matrix elements, which govern the quadratic Zeeman effect will have $J'=0$ or $J'=2$, thus the perturbing levels are {\it f} levels with positive parity. We note that a quadratic contribution in $B$ will also  occur for $\Delta J=0$ because diagonal Zeeman terms will transform in non-diagonal terms ($\Delta \Omega=+/-1$) in a molecular basis because $l$ and $j$ from the atom pair basis are no longer good quantum numbers for the molecular case. Levels with $J'=0$ and 2 can only occur for states $0^-, 1$ and $2$, where the first and last one correlate to the asymptotes $^3$P$_{0,2} + ^1$S$_0$ and thus only accidental coincidences with the observed levels will give significant contributions. But the levels $J'=2-$ of $\Omega=1$ correlating to the asymptote $^3$P$_1 + ^1$S$_0$ run parallel with the observed level structure and a regular pattern of the perturbation can be expected. We calculated these for the derived potential scheme and found that in all observed cases a level $J'=2-, \Omega=1$ is above the observed $J=1-$ level of $\Omega=0^+$ and below the one of $\Omega=1$. This leads to level shifts down in energy for the $0^+$ levels and up for $1$ and with the sign definition of the scalar susceptibility $\chi^{scal}$ in eq. \ref{Eq:tensor} to positive and negative values for $0^+$ and $1$, respectively. The slow decrease of the magnitude of  $\chi^{scal}$  for deeper bound levels corresponds to the increase of the level spacing going deeper into the potential. 

A similar calculation has been performed for Sr$_2$ \cite{mcg13}. For the case $0^+$ the authors approximate the quadratic Zeeman shift as being inversely proportional to the difference of the potential functions of $0^+$ and $1$. This is justified because as mentioned in the introduction the quadratic Zeeman part for the bound levels very closely lying to the asymptote is dominated by the coupling to the continuum. Consequently, this approach fails for the observation of the deeper bound levels of state $1$, because the perturbing level belongs to the same potential as the observed one and the energy difference originates from the rotational energy. This influence can be seen from significant deviations in Fig. 3 of reference \cite{mcg13}.

\section{conclusion}

In conclusion we have calculated the influence of the quadratic Zeeman effect on the photoassociation lines of $^{40}$Ca near the $^3$P$_1$ + $^1$S$_0$ asymptote and shown that it is much smaller than in the case of Sr as a result of the very weak dipole - dipole interaction in calcium and the dominating van der Waals interaction in long range. The previous measurements have been corrected for this effect since it was considerably larger than the estimated uncertainties for the unperturbed lines. Improved values of the long range parameters $C_6$ and $C_8$ have been derived that agree with the previously derived ones to a few percent and 20\%, respectively, well within the previously estimated uncertainties.    

\begin{acknowledgments}
This work was supported by Deutsche Forschungsgemeinschaft (DFG) through the {\it Center of Quantum Engineering and Space-Time Research (QUEST)} of the Leibniz Universit\"at Hannover and through the Research Training Group 1729 {\it Fundamentals and Applications of Ultra Cold Matter}. 
We thank Sebastian Kraft, Stefan Schulz and Oliver Appel for experimental support in the early stage of this work and Veit Dahlke for discussions.
\end{acknowledgments}

\bibstyle{pra}

\end{document}